\documentclass[12pt,preprint]{aastex}

\shorttitle{X-ray Sources around Alpha Cen}
\shortauthors{T.\ R.\ Ayres}
\begin{document}

\title{Serendipitous X-ray Sources in the \\ {\em Chandra}\/ HRC Field
around Alpha Centauri}

\author{Thomas R.\  Ayres}

\affil{Center for Astrophysics and Space Astronomy,\\
389~UCB, University of Colorado,
Boulder, CO 80309;\\ Thomas.Ayres@Colorado.edu}

\begin{abstract}
For more than a decade, Alpha Centauri AB (G2~V\,+\,K1~V) has been observed by {\em Chandra,}\/ in a long-term program to follow coronal ($T{\sim}10^{6}$~K) activity cycles of the two sunlike stars.  Over 2008.4--2017.8, nineteen HRC-I exposures were taken, each about 10~ks in duration, and spaced about six months apart.  Beyond monitoring the AB X-ray luminosities, the HRC-I sequence represents a unique decadal record of the dozen, or so, serendipitous X-ray sources in the $\alpha$~Cen field, which is at low Galactic latitude and thus dominated by nearby stars.  For the present study, the ten brightest candidates were considered.  Only a handful of these were persistent; most were variable, some highly so, flaring in a few epochs, weak or absent in the others.  All ten X-ray sources have {\em Gaia}\/ objects within about 2\arcsec; mostly late-type dwarfs, but a few giants.  However, two of the proposed optical counterparts have statistically significant offsets, and possible conflicts between the X-ray and optical properties.  Another of the candidates brightened a factor of 100 in X-rays during a single exposure, briefly attaining super-flare status.  The {\em Gaia}\/ counterpart is anomalously blue for its absolute $G$-magnitude, and likely is a WD\,+\,dM pair.  To the extent that the low Galactic latitude field is representative, the {\em Chandra}\/ time-domain view emphasizes that the high-energy stellar sky is biased toward transient sources, so any snapshot survey surely will miss many of the most interesting objects.
\end{abstract}

\keywords{ stars: coronae --- stars: individual (HD\,128620, 128621)  --- X-rays: stars}

\section{INTRODUCTION}

As one tunes up through the electromagnetic spectrum from the visible to the ultraviolet then X-rays, cosmic objects, especially late-type stars, tend to become progressively more variable; firmly entrenching the high-energy sky in the realm of time-domain astronomy.  Despite this, temporal studies are the exception for the most powerful contemporary X-ray observatories, {\em Chandra}\/ and {\em XMM-Newton}.\/   One reason is that these facilities are heavily over-subscribed: monitoring programs generally require a significant investment of time, often over multi-year periods, testing the enthusiasm of review panels.   A second reason is that observatory administrators tend to view time-domain projects as burdensome on their schedulers, especially in the case of {\em Chandra}\/ for which increasingly challenging thermal management of the spacecraft adds an additional layer of complexity for scheduling (consequently, programs with ``time constraints'' are discouraged).

There are notable exceptions to the general rule, mainly involving long-duration exposures of certain regions of the sky, for example the {\em Chandra}\/ Deep Fields (e.g., Giacconi et al.\ 2002); or the 1~Ms pointing on the Orion Nebular Cluster (so-called COUP survey: Sciortino et al.\ 2005).  As far as stellar astronomy is concerned, however, the CDF fields were extragalactic-dominated, as free as possible of ``contaminating'' foreground stars; while the COUP effort, although stellar oriented, focused on very young objects (Wolk et al.\ 2005).

The present study considers a third exception:  a {\em Chandra}\/ program, initiated in October 2005, to follow the X-ray histories of the two central stars -- ``A'' (G2~V) and ``B'' (K1~V) -- of the nearby Alpha Centauri triple system.  The motivation was to characterize the long-term coronal ($T{\sim}10^{6}$~K) behavior of these sunlike dwarfs, analogous to the 11-year periodicity of magnetic spots on our own Sun; in an effort to better understand the enigmatic cycle engine, the so-called Dynamo.  Because the ebb and flow of the stellar activity was anticipated to last many years, the program could make use of semi-annual, easy-to-schedule visits.  And, because the two very nearby stars (1.3~pc) are bright in soft X-rays, the program required only modest exposure times.  {\em Chandra,}\/ itself, proved indispensable to the effort, because in recent years AB passed through a close approach on the sky in their 80-year eccentric orbit: the tight separation (4\arcsec--6\arcsec\ for 2011--2021) has required {\em Chandra's}\/ excellent ($\sim$1\arcsec) spatial resolution to isolate the individual X-ray sources free of interference from the other.  A recent progress report on the $\alpha$~Cen coronal cycles can be found in Ayres (2018).

While the main goal of the $\alpha$~Cen program was to record the coronal luminosities of AB over time, there are a number of other X-ray sources in the field, whose emission histories also have been captured over the decade duration of the project.  These serendipitous sources are the subject of the present study: a glimpse into the type of objects, and their degree of variability, that might be caught in an arbitrary field of the Galactic plane (thus dominated by nearby stars) in a long-term, multi-epoch high-energy campaign, such as planned for the up-coming all-sky eROSITA mission ({\em e\,}xtended {\em RO\,}entgen {\em S\,}urvey with an {\em I\,}maging {\em T\,}elescope {\em A\,}rray: Predehl \& eROSITA Team 2013).

\section{OBSERVATIONS}

\subsection{{\em Chandra}\/ High-Resolution Camera}

{\em Chandra's}\/ High-Resolution Camera (HRC-I) was chosen for the $\alpha$~Cen project because the microchannel plate design is insensitive to ``optical loading'' (adverse impact of visible photons, also called ``red leak'') and ``pile-up'' (multiple high-energy events in a single pixel, relevant for slow-framing CCD-type X-ray detectors); two important considerations for observing visually bright stars that also are counts-per-second X-ray sources.  In addition, HRC-I has excellent low-energy response, a crucial advantage for sunlike soft coronal sources (AB have been recorded on several occasions by {\em Chandra's}\/ Low Energy Transmission Grating Spectrometer [LETGS], confirming that both spectral distributions are relatively cool, $T{\sim}$1--2~MK, even in their coronal high states:  Ayres 2014).  Further -- and relevant to the present study --  {\em Chandra's}\/ high spatial resolution minimizes source confusion and spreads out the diffuse cosmic background, aiding detection of faint objects.  

The main downside of HRC-I is that it has minimal spectral response:  it can provide a broad-band X-ray flux, but little information concerning the source temperature or any soft absorption.  That limitation was not a serious concern for the $\alpha$~Cen program, because the periodic LETGS exposures provided detailed energy spectra at different phases of the respective AB coronal cycles.  Lack of energy discrimination also was not a concern for the present project, because the other X-ray objects in the $\alpha$~Cen field are faint, and would not have provided useful spectra even in a long-duration pointing by, say, {\em Chandra's}\/ energy-resolving CCD imaging system (ACIS).  At the same time, the typically elevated X-ray/optical luminosity ratios of the serendipitous sources suggest that they all have high coronal temperatures, $\sim$10$^{7}$~K, in a regime where the conversion of count rates to energy fluxes is relatively independent of the source spectrum (although interstellar absorption can come into play for the more distant objects).

\subsection{Multi-Epoch HRC-I Time Series}

In the first few years of the {\em Chandra}\/ $\alpha$~Cen program, the pointings were relatively short, only 5~ks in duration.  Beginning in mid-2008, however, the exposures were lengthened to 10~ks, mainly to facilitate removal of transient flares, especially for the more active B component.  Recently, in mid-2018, the program returned to 5~ks pointings, because the few flares so far captured (exclusively from B) were brief enough that a quiescent X-ray level could be reliably estimated even in the shorter duration exposures.  

However, for the purpose of detecting and time-resolving serendipitous sources in the $\alpha$~Cen field, the deeper, higher-sensitivity  10~ks pointings are preferred.  There are nineteen of these, taken between 2008.4--2017.8, on a roughly semi-annual basis.  Table~1 summarizes the 10~ks exposures.  Included are astrometric offsets derived by comparing the X-ray center-of-mass coordinates of $\alpha$~Cen AB in each epoch with predictions based on the known proper motion and parallax of the binary.  The astrometric shifts rarely were larger than 0.5\arcsec\ in magnitude ($\sqrt{\Delta{x}^{2}\,+\,\Delta{y}^{2}}$, where $x$ and $y$ refer to the right-ascension-like and declination-like coordinates of the event lists, both in arcseconds), attesting to the excellent aspect reconstruction.  The empirical offsets, nevertheless, were applied to the individual event lists, for example when co-adding them to produce a time-averaged X-ray image in a fixed celestial reference frame, to avoid the -- albeit slight -- blurring of any point sources (at least for those with negligible proper motions over the 10 year period).

Figure~1, left panel, is a cumulative X-ray map of the inner 16\arcmin${\times}$16\arcmin\ of the $\alpha$~Cen HRC-I  field, where the apparent X-ray sources suffer less vignetting, and all are contained within the  boundaries of the collection of nineteen exposures in the face of the associated field rotations due to the different pointing orientations.  The map was constructed relative to a fixed center (219.8725\arcdeg, $-$60.8330\arcdeg), so high proper motion $\alpha$~Cen appears as a streak.  The latter is magnified in the inset panel, showing the westerly drift of the binary, but also the closing separation and rotating position angle of the two components owing to the evolving orbit (which passed through an apparent minimum in 2016), together with superimposed wiggles due to the annual parallax cycles.  In the main panel, the gray scale was set to emphasize positive detections at the lower intensities, but toned down in the center for $\alpha$~Cen AB, which are a hundred times, or more, brighter than the other sources.

Smaller circles mark entries from the {\em Chandra}\/ Source Catalog, Version 2\footnote{see: http://cxc.harvard.edu/csc/}:  blue were flagged in the catalog as positive detections; red, marginal.  Larger green numbered circles indicate the ten objects-of-interest considered in the present study.

Right hand panels of Fig.~1 depict the individual sources in each of the nineteen epochs.  The sub-images are 24\arcsec${\times}$24\arcsec, centered on the X-ray position (center-of-mass coordinates for $\alpha$~Cen AB).  Red outlined frames at far right are series averages (excluding AB).   Image scaling is the same for the ten objects-of-interest, but again toned down for AB.  It is clear from the figure that of the serendipitous sources, only 4, 6, and 10 are present in the majority of the exposures, although even then showing noticeable changes across the individual time series.  The other sources are more dramatically variable, conspicuous in only one, or a few, of the frames (especially Sources 2, 5, and 8).

\clearpage
\begin{deluxetable}{rcrcc}
\tabletypesize{\small}
\tablenum{1}
\tablecaption{{\em Chandra}\/ HRC-I 10~ks Observations of the $\alpha$~Cen Field}
\tablecolumns{5}
\tablewidth{0pt}
\tablehead{\colhead{ObsID} & \colhead{UT Start} & \colhead{$t_{\rm exp}$} &  \colhead{${\Delta}x$} &  \colhead{${\Delta}y$} \\
\colhead{} & \colhead{} & \colhead{(ks)} & \colhead{(\arcsec)} & \colhead{(\arcsec)}  \\
\colhead{(1)} & \colhead{(2)} & \colhead{(3)} & \colhead{(4)} & \colhead{(5)}    
} 
\startdata
 8906  &    2008.389  &    10.08  &  $+0.29$  &  $+0.10$  \\[1mm]
 8907  &    2008.961  &     9.34  &  $-0.25$  &  $-0.09$  \\[1mm]
 9949  &    2009.409  &    10.06  &  $+0.27$  &  $+0.22$  \\[1mm]
 9950  &    2009.949  &    10.05  &  $-0.28$  &  $-0.09$  \\[1mm]
10980  &    2010.335  &     9.76  &  $-0.14$  &  $+0.21$  \\[1mm]
10981  &    2010.808  &    10.03  &  $-0.21$  &  $-0.39$  \\[1mm]
12334  &    2011.993  &    10.07  &  $-0.06$  &  $-0.41$  \\[1mm]
14191  &    2012.473  &    10.10  &  $+0.20$  &  $+0.04$  \\[1mm]
14192  &    2012.950  &    10.06  &  $+0.09$  &  $-0.44$  \\[1mm]
14193  &    2013.480  &    10.59  &  $+0.16$  &  $+0.28$  \\[1mm]
14232  &    2013.963  &    10.05  &  $-0.47$  &  $-0.11$  \\[1mm]
14233  &    2014.477  &     9.62  &  $-0.04$  &  $+0.18$  \\[1mm]
14234  &    2014.999  &    10.11  &  $-0.13$  &  $-0.25$  \\[1mm]
16677  &    2015.346  &    10.07  &  $-0.15$  &  $+0.33$  \\[1mm]
16678  &    2015.810  &    10.08  &  $-0.26$  &  $-0.88$  \\[1mm]
16679  &    2016.330  &    10.03  &  $-0.15$  &  $+0.41$  \\[1mm]
16680  &    2016.771  &    10.01  &  $+0.52$  &  $+0.32$  \\[1mm]
16681  &    2017.335  &     9.99  &  $+0.36$  &  $+0.18$  \\[1mm]
16682  &    2017.823  &    10.00  &  $-0.02$  &  $-0.07$  \\[1mm]
\enddata
\tablecomments{Col.~3 is net exposure, corrected for dead time.  Cols.~4 and 5 are astrometric corrections.}
\end{deluxetable}

\clearpage
\begin{figure}[ht]
\figurenum{1}
\hskip  -15mm
\includegraphics[width=1.1\linewidth]{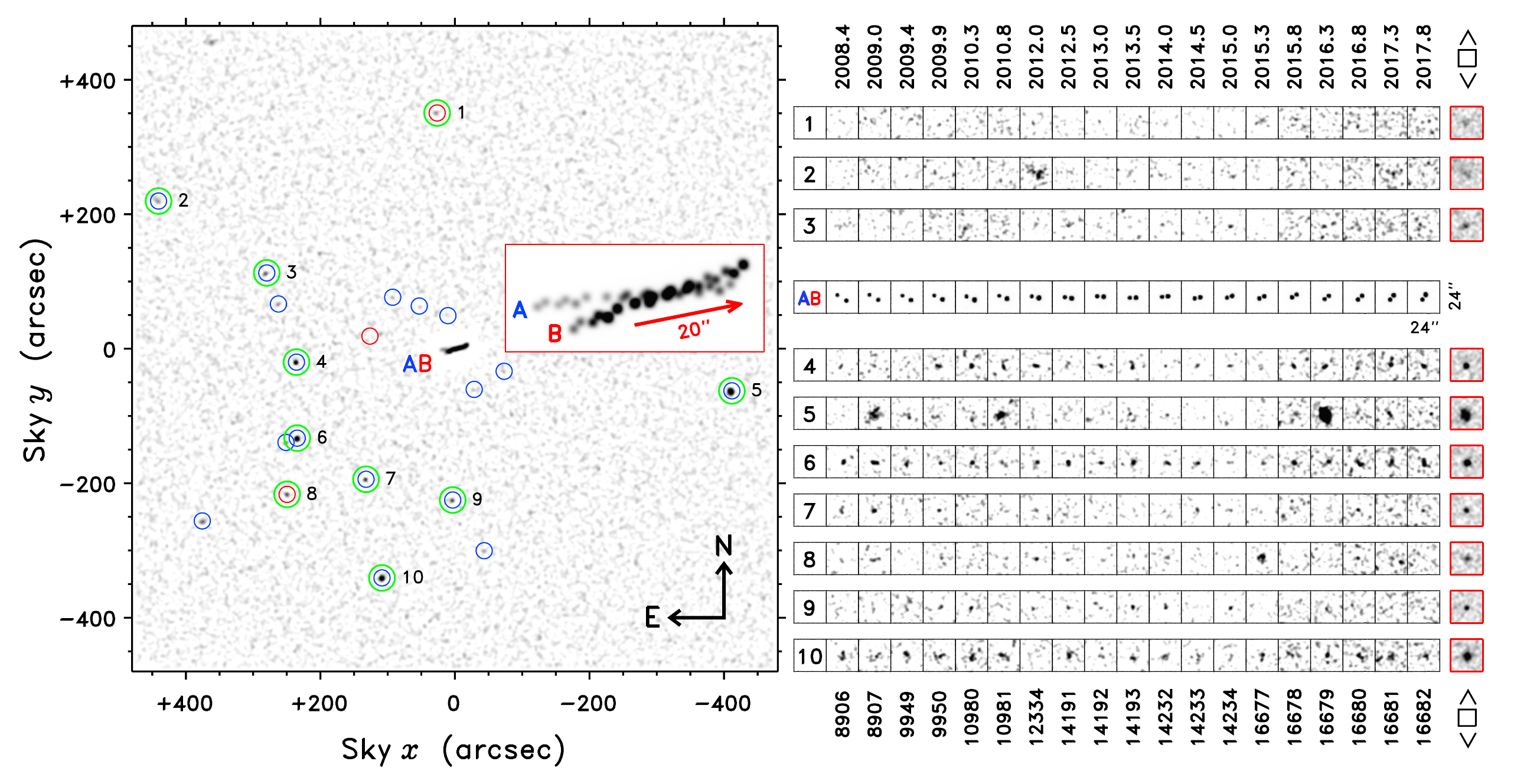} 
\vskip  -3mm
\caption[]{\small
Left hand panel is co-added X-ray map over the full 19 epochs of the $\alpha$~Cen ``10-ks'' sequence.  Inset shows a higher-resolution view of the $\alpha$~Cen streak image over the 10-year interval.  Red arrow indicates a displacement of 20\arcsec\ in the direction of the proper motion.  Right hand panels depict individual exposures of each of the ten objects-of-interest.  The gray scaling is more muted for the much brighter $\alpha$~Cen stars.}
\end{figure}
\clearpage
\begin{deluxetable}{ccccccccccc}
\rotate
\tabletypesize{\scriptsize}
\tablenum{2}
\tablecaption{{\em Chandra}\/ HRC-I X-ray Positional Measurements and {\em Gaia}\/ Counterparts}
\tablecolumns{11}
\tablewidth{0pt}
\tablehead{ \colhead{Target} & \colhead{RA} & \colhead{DEC} &\colhead{$\rho$} & \colhead{{\em Gaia}\/ DR2} & \colhead{$\varpi$} & \colhead{$G$} &\colhead{$(G_{BP}-G_{RP})$} & \colhead{$r_{\rm obs}$} & \colhead{$r_{\rm 95}$}  & \colhead{Detection $s$ [$N$]}   \\
\colhead{} & \colhead{($\arcdeg$)} & \colhead{($\arcdeg$)} & \colhead{($\arcmin$)} & \colhead{} & \colhead{(mas)} &\colhead{(mag)} &\colhead{(mag)} &\colhead{($\arcsec$)} &  \colhead{($\arcsec$)} & 
\colhead{}   \\
\colhead{(1)} & \colhead{(2)} & \colhead{(3)} & \colhead{(4)} & \colhead{(5)}  & \colhead{(6)}   & \colhead{(7)}   & \colhead{(8)}   & \colhead{(9)}  & \colhead{(10)} & \colhead{(11)}  
} 
\startdata
  1 & 219.887787 & $-60.735639$ & 5.9 & 5878499850885308928 & $1.74{\pm}0.03$ & +12.76 & +0.80 & 0.75 & 2.2 &  2.8~[4]   \\[2mm]
  2 & 220.123326 & $-60.771728$ & 8.2 & 5877749747781204096 & $3.80{\pm}0.15$ & +17.53 & +2.16 & 1.00 & 1.1 &  5.9~[4]   \\[2mm]
  3 & 220.032000 & $-60.801811$ & 5.0 & 5877748923147297536 & $0.31{\pm}0.03$ & +12.31 & +0.93 & 0.20 & 0.86 &  4.0~[5]   \\[2mm]
 4 & 220.006699 & $-60.838614$ & 3.9 & 5877747754933326208 & $0.45{\pm}0.20$ & +18.18 & \nodata  & 0.95 & 0.28 & 11~[16]\\[2mm]
  5 & 219.638437 & $-60.850511$ & 6.9 & 5878474772572315776 & $8.93{\pm}0.15$ & +17.81 & +1.72 & 0.75 & 0.31 & 19~~[3]   \\[2mm]
  6 & 220.005753 & $-60.870028$ & 4.5 & 5877747376976147456 & $0.88{\pm}0.03$ & +14.47 & +1.32 & 0.34 & 0.24 & 13~[16]   \\[2mm]
  7 & 219.947901 & $-60.886808$ & 3.9 & 5877724081049229312 & $0.00{\pm}0.13$ & +17.66 & +0.55 & 0.24 & 0.51 &  6.4~[8]   \\[2mm]
8 & 220.014262 & $-60.893075$ & 5.5 & 5877747273879148160 & $5.78{\pm}0.26$ & +18.60 & \nodata  & 0.64 & 0.88 &  7.0~[7] \\[2mm]
  9 & 219.874159 & $-60.895431$ & 3.7 & 5877724733882922368 & $0.41{\pm}0.03$ & +11.48 & +1.71 & 0.67 & 0.56 &  6.3~[7] \\[2mm]
10 & 219.933971 & $-60.927606$ & 6.0 & 5877723668732031616 & $0.61{\pm}0.42$ & +19.37 & \nodata  & 2.42 & 0.50 & 13~[14]\\[2mm]
\enddata
\vskip -3mm
\tablecomments{Col.~4 is radial displacement in arcminutes of target from reference coordinates (approximate average image center: 219.8725\arcdeg, $-$60.8330\arcdeg); radius of the detect cell is:  $(1.5 + 0.07\,\rho^{2.2})\arcsec$, with $\rho$(\arcmin).  Col.~9 is distance ($r= \sqrt{{\Delta}x^2\,+\,{\Delta}y^2}$) of target from nearest {\em Gaia}\/ DR2 entry.   Col.~10 is radius of X-ray error circle at 95\% confidence level.  Col.~11 is photometric significance of the detection, over the selected epochs, in one-sided Gaussian sigma; value in square brackets is number of exposures used for the positional measurement and to assess the detection significance.}
\end{deluxetable}

\clearpage
\section{ANALYSIS}

The following sections describe measurements of X-ray source positions and count rates.  A 95\% encircled energy detect cell radius, $1.5 + 0.07\,\rho^{2.2}$ arcseconds, was adopted, where $\rho$ is the radial displacement of the source from the image center in arcminutes, and the second term accounts for the blooming of the source size owing to vignetting.  The diffuse cosmic background was assayed in an annulus 50\arcsec--100\arcsec\ centered on each source in each exposure. 

\subsection{X-ray Source Positions}

X-ray positions of the 10 objects-of-interest were centroided in co-added event lists assembled from subsets of the epochs, restricted to those in which the source was clearly present: e.g., 2, 6, and 16 for Source~5.  Table 2 lists the derived coordinates.   The measured X-ray positions were matched against {\em Gaia}\/ Data Release 2 (DR2: Forveille et al.\ 2018)\footnote{see also: https://gea.esac.esa.int/archive/}.  Candidate optical counterparts, within about 2\arcsec, were found {for all ten objects.}  Of these, eight were considered good matches: namely, agreement with the DR2 coordinates within roughly the 95\% confidence X-ray error circle (i.e., the radius at which only one out of twenty centroiding trials would lead to a larger value by chance).  The error radii were set by Monte Carlo modeling of the centering process for the specific characteristics of each source (size of the off-axis detect circle, number of net source counts in the cell, and average areal density of the background counts: see Ayres 2004).  The typical calculated $r_{\rm 95}$ were less than 1\arcsec, except for one case, low signal-to-noise Source~1, where $r_{\rm 95}$ was closer to 2\arcsec. 

The other two cases were mismatches, at least given the criteria mentioned above: Source 4, with an empirical offset of 1.0\arcsec, about three times the 95\% error radius; and Source 10, which had a significantly larger offset, 2.4\arcsec, from the nearest {\em Gaia}\/ object, compared to the predicted 95\% radius of only 0.5\arcsec.  Under practically any other circumstances, these arcsecond-scale offsets would be considered acceptable matches, but the high quality of both the HRC-I X-ray centroids and the {\em Gaia}\/ optical coordinates make the statistical deviations hard to ignore.  These mismatches will be discussed in greater depth later.

{{\em Gaia}\/ DR2 was essential to this study because the normal catalog-of-choice for optical counterparts, U.S. Naval Observatory B1.0 (Monet et al.\ 2003), was entirely devoid of matches to the $\alpha$~Cen serendipitous X-ray sources.  This is because photographic-based B1.0 is strongly affected in the vicinity of bright stars by their heavily saturated images on the original Schmidt sky-survey plates, so that fainter stars were mostly washed out.  The 2MASS infrared sky survey (Skrutskie et al.\ 2006), which utilized digital cameras, is less affected by bright-star saturation, but has relatively shallow flux limits compared to {\em Gaia}\/ or USNO B1.0.  Consequently, only a few of the typically optically faint $\alpha$~Cen serendipitous X-ray sources have 2MASS counterparts.} 

The positional agreements (or disagreements) were based solely on the estimated random errors associated with the Poisson measurement process, and neglected any systematic errors that potentially could influence the outcomes.  These might include subtle geometrical distortions of the HRC-I field-of-view, especially far from the boresight (where most of the objects-of-interest fall); or proper motions of the targets themselves.  However, the fact that eight of the ten candidates show reasonably good agreement with the {\em Gaia}\/ astrometry, at the sub-arcsecond level, suggests that the first possibility probably is not a factor.  Further, the modern epoch of the {\em Gaia}\/ survey (2015.5 for DR2, close to the 2013.1 mid-point of the $\alpha$~Cen 10~ks series) mitigates against the second possibility.  {In fact, the proper motions reported for Sources~4 and 10 in DR2 are an insignificant few mas yr$^{-1}$.}

\subsection{X-ray Count Rates}

Count rates of the ten objects-of-interest were measured in the nineteen individual exposures, as well as in the sum.  If there were sufficient source counts in an epoch, the measurements were made in sliding temporal windows designed to achieve a fixed detection significance in each sub-interval (usually about 20 counts total, depending on the background).  An exception was $\alpha$~Cen: bins were sized to accumulate 1500 cnt for A and 4000 cnt for (normally brighter) B.  Another exception was epoch 16 (2016.3) of Source 5, when the object experienced a large X-ray outburst: the sliding bins were set at 2~ks during the quiescent period in the first half of the exposure; decreasing to 1~ks at the beginning of the outburst; then 0.5~ks during the rapid flare rise and subsequent decay. 

The resulting light curves are illustrated in Figure~2.  Note the scale change for $\alpha$~Cen AB.  Most of the objects were faint enough during the individual epochs that only the average count rate is displayed.  The few cases where the sliding-bin measurements were triggered are marked in red.  Green dots at the right side of the main panel depict count rates averaged over the full time series.  Note for Source~5 the apparent flare decays in epochs 2 and 6, and the large outburst in 16.  The light curve of the latter is shown separately to the right of the main panel.  In the other intervals, Source 5  was much weaker or not detected at all.  There also were a few conspicuous flares in the $\alpha$~Cen B time series, especially at the peak of its activity cycle in exposures 5--7; but nothing so obvious in the $\alpha$~Cen A light curves.
  
\clearpage
\begin{figure}[ht]
\figurenum{2}
\hskip  7mm
\includegraphics[width=0.9\linewidth]{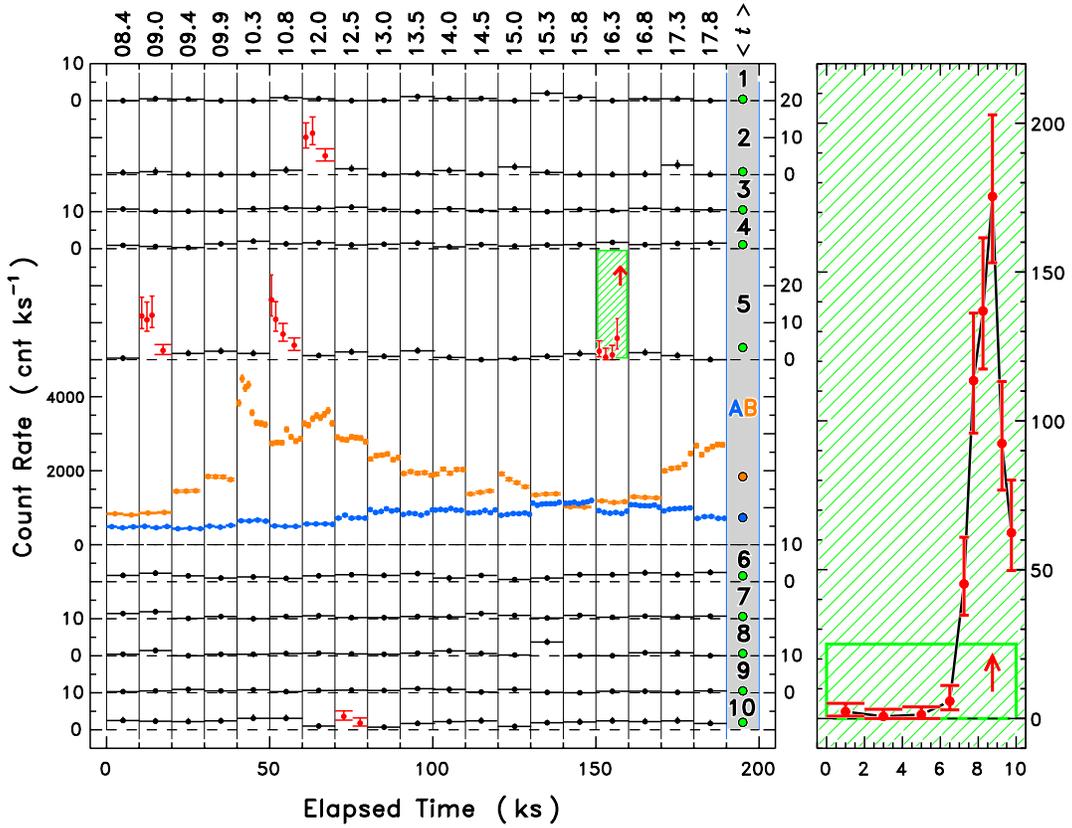} 
\vskip  -7mm
\caption[]{\small
X-ray light curves for the ten objects-of-interest and $\alpha$~Cen AB.  Note the $y$-axis scale change for the latter.  Horizontal bars on the count-rate points indicate durations of the individual time bins (normally, just the full 10~ks of the exposure).  Vertical error bars, where visible, are 90\% confidence intervals.  Red points are for epochs where high enough count rates permitted finer division of the time series.  Light curve of the large X-ray flare of Source~5 in exposure 16 (green-hatched sub-frame: flare peak off-scale) is shown on an expanded scale at right.}
\end{figure}

Table 3 lists count rates averaged over the full time series; the highest average epoch count rate in the sample of 19; and the maximum count rate encountered in any sliding bins within individual exposures.  Large deviations of the latter two measures from the campaign averages highlight sources that are strongly variable.

\clearpage
\begin{deluxetable}{ccccccccccc}
\rotate
\tabletypesize{\scriptsize}
\tablenum{3}
\tablecaption{Source Properties: Bolometric and X-ray Fluxes, Luminosities, and Ratios}
\tablecolumns{11}
\tablewidth{0pt}
\tablehead{ \colhead{Target} & \colhead{Sp.~Typ.} & \colhead{(B.C.)$_{G}$} & \colhead{$f_{\rm bol}$} &  \colhead{CR$_{\rm ave}$} &  \colhead{CR$_{\rm epoch}$} &  \colhead{CR$_{\rm max}$} & \colhead{ECF} & \colhead{$f_{\rm X}$} & \colhead{$\log{L_{\rm X}}$} & \colhead{$\log{L_{\rm X}/L_{\rm bol}}$}    \\
\colhead{} & \colhead{} & \colhead{(mag)} & \colhead{(cgs)} & \multicolumn{3}{c}{---~~~(cnt ks$^{-1}$)~~~---} & \colhead{} & \colhead{(cgs)} &\colhead{(cgs)} & \colhead{}   \\
\colhead{(1)} & \colhead{(2)} & \colhead{(3)} & \colhead{(4)} & \colhead{(5)}  & \colhead{(6)}   & \colhead{(7)}   & \colhead{(8)}   &  \colhead{(9)}   & \colhead{(10)} & \colhead{(11)}   
} 
\startdata
  1 & G V   & $+0.1$ &   206 &   $0.4_{-0.1}^{+0.1}$   &  2.0 &      2.0  & 1 &  0.4    &    +29.20 & $-4.71$  \\[3mm]
  2 & M V   & $-0.8$ &     5.8 &   $0.8_{-0.2}^{+0.2}$   &  7.2 &   11.2   & 1 &  0.8    &    +28.82 & $-2.86$  \\[3mm]
  3 & G III  & $+0.1$ &    880 &   $0.5_{-0.1}^{+0.1}$   &  1.2 &     1.2   & 2 &  1.0    &    +31.10 & $-4.94$  \\[3mm]
  4 & K V:   & $-0.1$ &     4.7 &   $1.1_{-0.1}^{+0.1}$   &  2.0 &     2.0   & 2 &   2.2   &    +31.12 & $-2.33$  \\[3mm]
  5 & M V + WD   & $-1.7$ &  10.3 &   $3.3_{-0.2}^{+0.2}$   &  33 &     175   & 1 &   3.3   &    +28.70 & $-2.50$  \\[3mm]
  6 & K IV   & $-0.2$ &  160 &   $1.6_{-0.2}^{+0.1}$   &  2.5 &     2.5   & 2 &   3.2   &    +30.70 & $-3.69$  \\[3mm]
  7 & F/G V?   & $+0.1$ &  6.3 &  $0.6_{-0.1}^{+0.1}$    &  1.9 &     1.9   & 2 &   1.2   &    +30.95 & $-2.72$  \\[3mm]
  8 & M V   & $-1.7$ &     5.0 &  $0.6_{-0.1}^{+0.1}$    &  3.7 &     7.2   & 1 &   0.6   &    +28.33 & $-2.92$  \\[3mm]
  9 & K III   & $-0.5$ &   3270 &  $0.5_{-0.1}^{+0.1}$    &  1.1 &     1.1   & 2 &   1.0   &    +30.85 & $-5.51$  \\[3mm]
10 & M V:  & $-0.8$ &    3.0 &  $2.0_{-0.2}^{+0.2}$    &  3.1 &     3.6   & 2 &   4.0   &    +31.11 & $-1.88$  \\ [3mm]
\enddata
\vskip -3mm
\tablecomments{Col.~4 units: $10^{-12}$ erg cm$^{-2}$ s$^{-1}$.  Col.~5 is average count rate over all 19 exposures; upper and lower bounds delimit 90\% confidence intervals.  These, and subsequent, count rates are for a 95\% encircled energy detect cell.  Col.~6 is maximum of the single-epoch averages.  Col.~7 is the maximum local count rate encountered in the whole time series (usually in a ``sliding-bin'' time interval).  Col.~8 Energy Conversion Factors are in $10^{-14}$ erg cm$^{-2}$ s$^{-1}$ (cnt ks$^{-1}$)$^{-1}$.  Col.~9 units: $10^{-14}$ erg cm$^{-2}$ s$^{-1}$.  Col.~10 units: erg s$^{-1}$.  Sources~4 and 10 have small parallaxes with large errors: the reported $\varpi$ was adopted to calculate the distance, acknowledging that the derived quantities ($L_{\rm X}$ and absolute magnitude $M_{G}$) are correspondingly uncertain (these also were the two sources with the poorest positional matches to {\em Gaia}\/ objects).  Source~7 had a zero reported parallax: distance was based on $\varpi= 0.4$~mas, $3\times$ the cited error.}
\end{deluxetable}

\clearpage
\subsection{Bolometric and X-ray Fluxes}

Table~3 also lists several derived quantities: bolometric fluxes at Earth, based on the {\em Gaia}\/ magnitudes and color-dependent bolometric corrections, and including an approximate correction for reddening; X-ray fluxes (again at Earth), based on the average HRC-I count rates, and also with an approximate correction for the interstellar hydrogen column; X-ray luminosities, based on the {\em Gaia}\/ distances; and X-ray-to-bolometric luminosity ratios (independent of distance and relatively insensitive to reddening).  

The bolometric flux transformation was derived from parameters published in Casagrande \& VandenBerg (2018); namely $G_{\odot}= -26.90$ and $({\rm B.C.})_{G, \odot}= +0.08$, the latter taken from their tables for solar metallicity and solar gravity at the solar effective temperature, to be consistent with the B.C.s derived later for the other stars (note also: $(G_{\rm BP}-G_{\rm RP})_{\odot}= +0.82$).  Given the apparent $G$-magnitude of the Sun, its distance, the bolometric correction, and the solar luminosity ($L_{\odot}= 3.83{\times}10^{33}$ erg s$^{-1}$: e.g., Ayres et al.\ 2006), one infers,
\begin{equation}
f_{\rm bol}= 2.55{\times}10^{-5}\,{\times}\,10^{-(G_{\star} + ({\rm B.C.})_{\star})\,/\,2.5}~~{\rm erg} ~{\rm cm}^{-2}~ {\rm s}^{-1}~~, 
\end{equation}
where $G_{\star}$ is the de-reddened {\em Gaia}\/ magnitude of the star of interest, and  $({\rm B.C.})_{\star}$ is the bolometric correction (Casagrande \& VandenBerg 2018).  

The {\em Gaia}\/ transformation is analogous to that for the Johnson $V$ band (see, e.g., Ayres 2017, Appendix A, eq.~1), and the numerical coefficient is nearly identical (not an accident, but rather a consequence of how the {\em Gaia}\/ $G$-magnitude was defined).  In contrast to $V$, however, the $G$-based transformation is somewhat less sensitive to reddening (see below), and the bolometric corrections are less strongly dependent on $T_{\rm eff}$, at least until one reaches the cooler M-types.  

For the X-rays, an Energy Conversion Factor (ECF) was applied to transform the HRC-I count rates into fluxes at Earth.  The ECF often is constructed to account for interstellar attenuation, so one obtains the ``un-absorbed'' flux at Earth, i.e., that which would be recorded in the absence of interstellar absorption.  The serendipitous sources in the $\alpha$~Cen field fall into two general categories: (1) those that have well-determined parallaxes, and are relatively nearby with distances $<$1~kpc; and (2) those that have poorer quality, or indeterminate, parallaxes, indicating larger distances, $>$1~kpc.  Given that this is an exploratory study, ECFs were fashioned for two scenarios: (1) nearby objects with modest ISM columns ($N_{\rm H}\sim 3{\times}10^{20}$~cm$^{-2}$; corresponding to a reddening of $E(B-V){\sim} 0.05$~mag); and (2) more distant objects with larger columns ($N_{\rm H}\sim 3{\times}10^{21}$~cm$^{-2}$; $E(B-V){\sim} 0.5$~mag).  These reddening values are roughly appropriate to the extinction maps for the Galactic plane published by Lallement et al.\ (2018), in the direction of $\alpha$~Cen and for the indicated distance ranges.  WebPIMMS\footnote{see: http://cxc.harvard.edu/toolkit/pimms.jsp}  simulations for the alternate columns, assuming a solar abundance, $T=10^{7}$~K APEC spectrum (and accounting for the 95\% encircled energy factor), yielded ECFs (for the 0.2--2~keV band) of $1{\times}10^{-14}$ and $2{\times}10^{-14}$ erg cm$^{-2}$ s$^{-1}$ (cnt ks$^{-1}$)$^{-1}$, respectively, for the un-absorbed fluxes.

The bolometric fluxes in the table were calculated from de-reddened $G$-band magnitudes for the alternate $E(B-V)$ values.  The absorption corrections were given by $A_{G}\sim 2.5\,E(B-V)$.  The latter was obtained by combining $A_{G}\sim 0.8\,A_{V}$ (last term is the total extinction for Johnson $V$) from a formula published by {\em Gaia}\/ Collaboration et al.\ (2018), together with the nominal Galactic extinction law $A_{V}\sim 3.1\,E(B-V)$.  The derived values were $A_{G}\sim$ 0.13 mag and 1.25 mag, respectively.  Also, reddening of the $(G_{BP}-G_{RP})$ color was compensated according to the approximate relation $E(G_{BP}-G_{RP})\sim 1.4\,E(B-V)$, as inferred from the selective absorption relations for $G_{BP}$ and $G_{RP}$ separately (ibid).  Note that the $G_{BP}$ and $G_{RP}$ band centers are further apart in wavelength than $B$ and $V$, thus the {\em Gaia}\/ color is more sensitive to the differential extinction.

The X-ray to bolometric luminosity ratio, $L_{\rm X}/L_{\rm bol}$, which is equivalent to the ratio of the apparent fluxes, $f_{\rm X}/f_{\rm bol}$, is invariant with distance, and roughly independent of reddening because the X-ray and optical extinction corrections described above are similar as a function of the interstellar column.  In contrast, the X-ray luminosity and absolute $G$-magnitude depend on distance (squared), so these parameters become more uncertain for the objects with poorly determined parallaxes.

Figure~3 illustrates a {\em Gaia}\/ H--R diagram for the ten objects-of-interest, reddening-corrected as described above; together with representative late-type (F--L) dwarfs, white dwarfs, and WD\,+\,dM binaries (significance of latter will become apparent shortly).  The context objects were collected from a variety of catalogs and source lists, restricted to nearby (minimally reddened), ostensibly single stars (except for the WD\,+\,dM systems).   The {\em Gaia}\/ diagram displays a tight, sloping late-type Main sequence (MS), as well as the parallel WD cooling sequence, shifted dramatically downward, and seemingly extended to even lower luminosities by the blue L dwarfs.  Meanwhile, there is a gap-bridging cloud of WD\,+\,dM binaries connecting the WD and normal-star tracks, as well as a lower (likely unrelated) bridge across the L dwarfs.

\clearpage
\begin{figure}[ht]
\figurenum{3}
\vskip     0mm
\hskip    5mm
\includegraphics[width=0.8\linewidth]{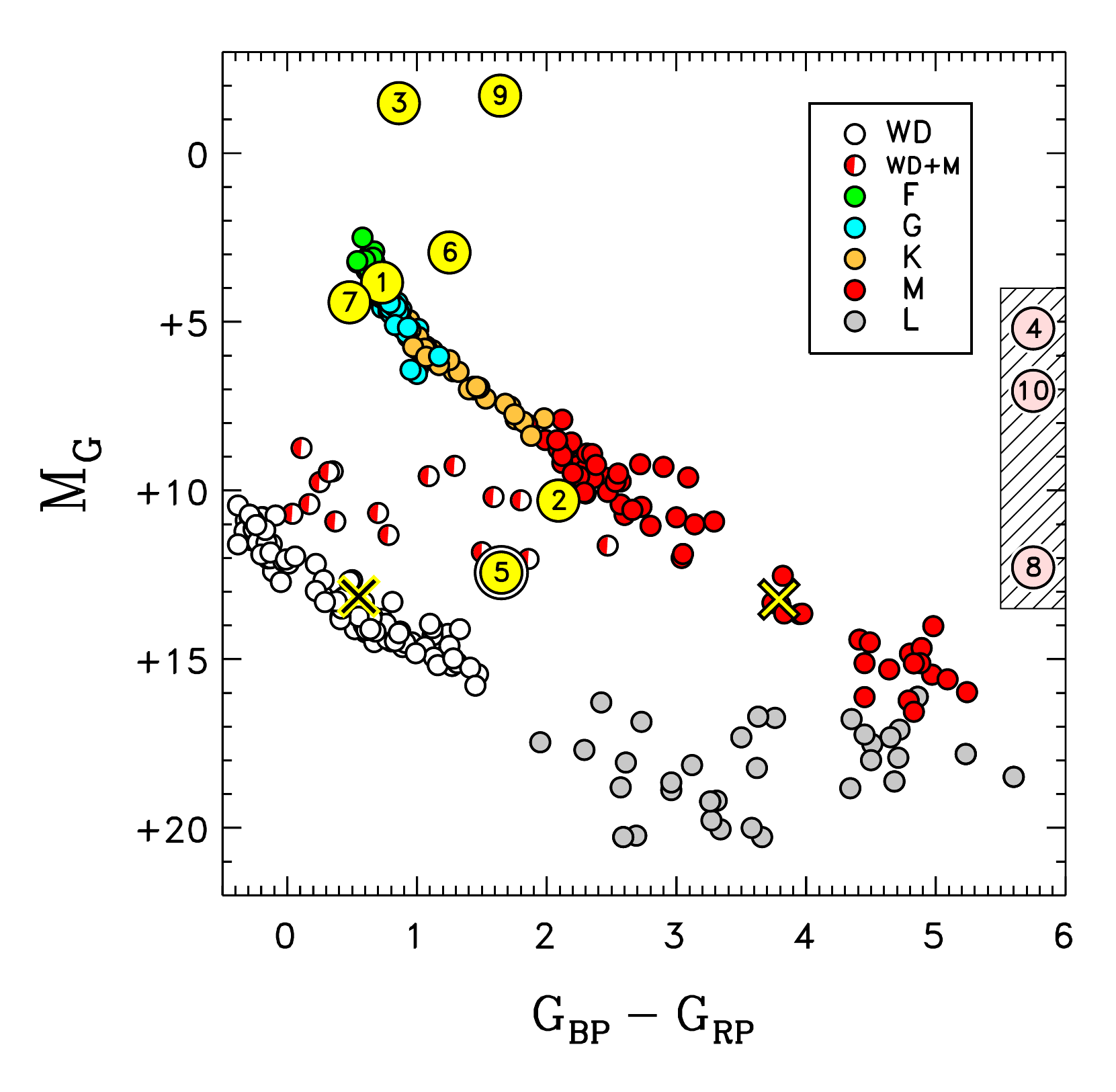} 
\caption[]{\small
{\em Gaia}\/ H--R diagram for representative single late-type (F--L) dwarfs, white dwarfs, and WD\,+\,dM binaries; together with the objects-of-interest (numbered 1--10) from the present program.  The three symbols in the hatched rectangle at right represent optical counterparts that lack colors in {\em Gaia}\/ DR2, but have at least approximate parallaxes to obtain the absolute $G$-magnitudes ($M_{G}$).  Source~5 has too blue a {\em Gaia}\/ color for its absolute $G$-magnitude, but falls in a region of the H--R diagram where WD\,+\,dM binaries are common: a pair of crosses depicts a decomposition of Source~5 into WD (left) and dM (right) components.}
\end{figure}

\clearpage
\section{DISCUSSION}

\subsection{The Individual Sources}

The serendipitous X-ray sources in the $\alpha$~Cen field can be characterized on several levels: optical properties (i.e., H--R diagram location); X-ray luminosity, $L_{\rm X}$; $L_{\rm X}/L_{\rm bol}$ ratio; and high-energy variability, especially flaring.  Also, because the $\alpha$~Cen field is projected against the high-extinction Galactic plane ({\em b}\,$= -0.7\arcdeg$), not far from the Galactic center, one can safely eliminate the possibility of contaminating extragalactic sources; a somewhat ironic reversal with regard to the {\em Chandra}\/ Deep Fields mentioned earlier.

\noindent
{\bf\em Source 1}\/  is the faintest X-ray star considered here, but the one with the most straightforward identification.  There is good coordinate agreement with a {\em Gaia}\/ object, which has a high-quality parallax.  {There also is a 2MASS entry (J14393306$-$6044075).}  The $G$-band absolute magnitude and {\em Gaia}\/ color point to a G dwarf.  {The 2MASS $(J-K)$ color is consistent with that choice.}  The X-ray luminosity (${\sim}10^{29}$ erg s$^{-1}$) and $L_{\rm X}/L_{\rm bol}$ ratio ($2{\times}10^{-5}$) indicate a moderately active star of young age, similar or younger than the Hyades (${\sim}$600~Myr).  The series-average X-ray source has a relatively low significance; most of the individual epochs have null count rates, but a few are noticeably elevated.

\noindent
{\bf\em Source 2}\/  also has a good {\em Gaia}\/ match, with a high-quality parallax.  The $G$-band absolute magnitude is consistent with an early-M dwarf, although the {\em Gaia}\/ color is slightly blueward of the MS.  However, the optical object is faint, so the $G_{BP}$ part of the color is less certain.  The source is highly variable in X-rays, dominated by an isolated brightening in epoch 7.  The average X-ray luminosity ($6{\times}10^{28}$) is consistent with an active M dwarf (e.g., Schmitt \& Liefke 2004), and the $L_{\rm X}/L_{\rm bol}$ ratio is in the ``saturated'' regime ($\sim 1{\times}10^{-3}$; as described, for example, by Wright et al.\ [2011] based on a large sample of late-type stars with both X-ray measurements and rotation periods: very few of their fast-rotating objects exceeded the saturation level by a factor of more than a few).   The saturated $L_{\rm X}/L_{\rm bol}$ and flaring behavior both are signatures of a hyperactive red dwarf.

\noindent
{\bf\em Source 3}\/  has excellent agreement with a {\em Gaia}\/ entry, and the $\sim$3~kpc distance has only a 10\% parallax error.  The absolute magnitude points to a class-III giant; the color, to a late-G star.  The X-ray luminosity is high (${\sim}10^{31}$), beyond the realm of the saturated dwarfs; but the $L_{\rm X}/L_{\rm bol}$ ratio is moderate (${\sim}10^{-5}$).  Both are symptoms of an active late-type giant, similar to fast-rotating HR\,9024 (Ayres et al.\ 1998).  Source~3 appears to be variable, although like Source~1, its overall significance in the full time series is relatively low, with only a few clear detections in the individual exposures.  {This object has the highest {\em Gaia}\/ proper motion for its parallax, of all the candidates with significant parallaxes, amounting to a tangential space velocity of about 100~km s$^{-1}$.}

\noindent
{\bf\em Source 4}\/  has a relatively high detection significance in the series average, and is one of the more persistent of the ten objects-of-interest, appearing in at least 16 of the 19 exposures.  Source 4 also is only about 4\arcmin\ from the image center, thus can be centroided more precisely than sources located further away.  Although there is a {\em Gaia}\/ object within 1\arcsec\ of Source 4, as explained earlier the positional mismatch is statistically large enough to question the identification.  Nevertheless, if the {\em Gaia}\/ object is the optical counterpart, the absolute $G$-magnitude would be consistent with a late-G or early-K dwarf, although noting that the cited parallax error is large (50\%).  Unfortunately, the {\em Gaia}\/ object is faint enough to lack a reported color.  The latter could strengthen, or refute, the identification.  Even so, the X-ray luminosity would be abnormally high for a late-type dwarf (${\sim}10^{31}$), more like a subgiant or giant (e.g., Source 3).   Accompanying the over-large $L_{\rm X}$ is an equally elevated $L_{\rm X}/L_{\rm bol}$ ratio, $\sim 5{\times}10^{-3}$, higher than the saturation limit seen among the most active G/K dwarfs.  Only a handful of the hyper-active members of the Wright et al. (2016) sample violated the $10^{-3}$ saturation ceiling, and then by less than a factor of 3.  On the other hand, a study of the very young ($\sim$13~Myr) h~Persei cluster by Argiroffi et al.\ (2016) found a number of pre-Main sequence members that were ``super-saturated,'' with $L_{\rm X}\lesssim 10^{31}$ erg s$^{-1}$ and $L_{\rm X}/L_{\rm bol}$ approaching $10^{-2}$; some, but not all, were binaries.  The conclusion is that it is possible -- but with serious caveats -- that the {\em Gaia}\/ object is responsible for Source~4; and, if it is, then the object is anomalously active, encroaching on the super-saturation regime.  The obvious quandary is that if the {\em Gaia}\/ object is not the match, then the true optical counterpart must be fainter still.  Even significantly more reddening likely would not modify the $L_{\rm X}/ L_{\rm bol}$ ratio enough to bring it down to a palatable level.  At the same time, Source~4 is one of the X-ray steadiest of the ten objects-of-interest, which otherwise would point to a normal active G/K coronal source.  In short, the identification of Source~4 presents a dilemma.

\noindent
{\bf\em Source 5}\/  is perhaps the most intriguing of the ten.  It is highly variable, with conspicuous flare decays in epochs 2 and 6, then a giant outburst in 16.  The rise and fall of the large flare were captured in the final 4~ks of that pointing, as illustrated earlier in Fig.~2.  {A possible} {\em Gaia}\/ counterpart {is} within 0.8\arcsec, although the 95\% confidence error radius is smaller by a factor of two.  {There is no entry in the main 2MASS catalog, but a match can be found in the Point Source Reject Table.}  The {\em Gaia}\/ object is relatively nearby, at about 110~pc, and has the absolute magnitude of a late-M dwarf, consistent with the flare-dominated behavior.  {The 2MASS $(J-K)$ color is red, as expected for an M dwarf.}  {At the {\em Gaia}\/ } distance, the series-average X-ray luminosity (including the flares) is $5{\times}10^{28}$, at the high end for active dM stars in the solar neighborhood (Schmitt \& Liefke 2004); and the $L_{\rm X}/L_{\rm bol}$ ratio, at about $4{\times}10^{-3}$, hovers at the top of the saturation ceiling (although both metrics are strongly biased by the large flare event).  All these signs point to an active late-M dwarf in the saturation regime.  The one oddity is the {\em Gaia}\/ color, which is too blue for a late-M dwarf.  However, as noted earlier, there is a horizontal band of WD\,+\,dM binaries that crosses between the WD cooling sequence and the lower part of the normal MS.  Source~5 sits at the bottom edge of the WD\,+\,dM bridge.  Given the tight relationships between absolute magnitude $M_{G}$ and $(G_{BP}-G_{RP})$ color in both the WD and dM empirical sequences, it is a simple matter to determine the specific WD\,+\,dM combination that satisfies the de-reddened absolute magnitude and color of Source~5.  The result is a WD with $M_{G}\sim +13.14$~mag and $(G_{BP}-G_{RP})\sim +0.55$~mag; and a red dwarf with $M_{G}\sim +13.23$ and $(G_{BP}-G_{RP})\sim +3.79$.  That pair is marked by crosses in Fig.~3.  

As for the large flare, the X-ray luminosity peaked at about $3{\times}10^{30}$ ergs s$^{-1}$, with a duration of perhaps 5~ks.  The X-ray maximum was a remarkable 20\% of the bolometric luminosity of the star.  The total radiated energy (0.2--2~keV), for the roughly triangular flare profile, approached $10^{34}$ erg.  This ranks the event as a ``super-flare'' (Maehara et al.\ 2012); more common in the vast {\em Kepler}\/ optical time-domain survey, but a rarer occurrence in the high-energy sky (because of observational bias; although {\em Swift's}\/ panoramic hard X-ray monitors have caught a few of the more energetic events: e.g., Osten et al.\ 2010).

\noindent
{\bf\em Source 6}\/  is perhaps the most constant of the ten in X-rays.  There is good coordinate agreement with a {\em Gaia}\/ entry, and the parallax of the counterpart is well determined.  The absolute $G$-magnitude and color suggest a K-type subgiant.  The high X-ray luminosity ($5{\times}10^{30}$) and moderately-high $L_{\rm X}/L_{\rm bol}$ ratio ($6{\times}10^{-4}$) point to an analog of the famous hyper-active RS~CVn systems like HR\,1099 (Walter et al.\ 1980).  These evolved, tidally synchronized binaries typically contain a late-type subgiant, often early-K, together with a MS star, often early/mid-G, in a few-day orbit.  Even though Source~6 is present in nearly all the exposures, it does display what appears to be a multi-year X-ray variation.

\noindent
{\bf\em Source 7}\/  is moderately variable in X-rays, appearing clearly in perhaps only half the epochs.  The X-ray position is a good match to a {\em Gaia}\/ object.  The optical counterpart has the bluest color of the ten, indicating an F or early-G spectral type.  Unfortunately, the {\em Gaia}\/ DR2 parallax is indeterminate.  Using the 3\,$\sigma$ parallax error as a limiting value, yields an absolute $G$-magnitude consistent with a MS star.  However, both the X-ray luminosity ($\sim 10^{31}$) and $L_{\rm X}/L_{\rm bol}$ ratio ($2{\times}10^{-3}$) are higher than for the most active single F/G dwarfs, more reminiscent of the PMS behavior found by Argiroffi et al.\ (2016).  Although the X-ray luminosity would change with any changes in the distance, the high $L_{\rm X}/L_{\rm bol}$ ratio would not, remaining uncomfortably above the saturation regime for single F/G stars.  More likely, the object is a tidally-synchronized binary of, say, late-F dwarfs at a somewhat greater distance than assumed, perhaps like the short-period $\sigma^2$~Coronae Borealis system (Osten et al.\  2003).  Mainly because of the unknown distance, the binary interpretation must be considered tentative (but, the most viable alternative).

\noindent
{\bf\em Source 8}\/ is similar to Source~2: weak or absent most of the time, but with a burst of activity in a single exposure (14).  The coordinates agree with a {\em Gaia}\/ entry; the parallax is well determined and large ($d\sim$170~pc).  The $G$-band absolute magnitude suggests a late-M, like the proposed red dwarf counterpart of Source~5.  Unfortunately, a confirming {\em Gaia}\/ color is lacking.  The  $L_{\rm X}$ and  $L_{\rm X}/L_{\rm bol}$ values are similar, though lower, than those of Sources~2 and 5, but still in the saturated regime for red dwarfs. 

\noindent
{\bf\em Source 9}\/ is relatively faint, and not as highly variable as Sources 2, 5, and 8.  There is a good fit to a {\em Gaia}\/ object, and the parallax is well characterized.  {There also is a 2MASS entry (J14392983$-$6053441).}  The optical counterpart has the absolute $G$-magnitude of a giant (like Source 3), and the $(G_{BP}-G_{RP})$ color of a K star, {also consistent with the red $(J-K)$ from 2MASS.}  The average $L_{\rm X}$ is high ($7{\times}10^{30}$), but the $L_{\rm X}/L_{\rm bol}$ ratio is moderate ($3{\times}10^{-6}$); both signs of a coronally active evolved star, although somewhat less extreme than HR\,9024, more like the K0 clump giant $\beta$~Ceti (Ayres et al.\ 1998). 

\noindent
{\bf\em Source 10}\/ is one of the problematic cases, almost identical to previously discussed Source~4.  Source~10 is more discrepant, however, with an offset relative to the nearest {\em Gaia}\/ object more than twice as large (2.4\arcsec).  Source~10 appears in at least 14 of the 19 exposures, showing a higher level of persistence than most of the other objects-of-interest, although some epoch-to-epoch variability also is present.  Like Source~4, the {\em Gaia}\/ optical counterpart is faint, near the survey limit; has a poor-quality parallax; and is missing a {\em Gaia}\/ color.  Taking the, albeit poorly-matched, {\em Gaia}\/ object as the true counterpart, the absolute $G$-magnitude would correspond to a mid-K dwarf, although again a confirming color is lacking.  Like Source~4, the series-average X-ray luminosity of Source~10 is high ($\sim 10^{31}$) and the $L_{\rm X}/L_{\rm bol}$ ratio ($1{\times}10^{-2}$) is extreme even for the super-saturated PMS stars described earlier.  If the true optical counterpart is unseen, below the survey limit for {\em Gaia}\/ DR2, then the $L_{\rm X}/L_{\rm bol}$ ratio would become implausibly large, even with substantially larger reddening than assumed for the $d>1$~kpc scenario (noting, again, that the 0.2--2~keV X-ray and {\em Gaia}\/ $G$-band attenuations behave quantitatively alike with the interstellar column).  As with Source~4, the strikingly similar Source~10 presents a dilemma.

\subsection{Identification Summary}

Eight of the ten X-ray objects-of-interest have reasonably good coordinate agreement with {\em Gaia}\/ entries.  The two objects with poorer {\em Gaia}\/ positional matches (Sources~4 and 10) also would challenge the X-ray/optical properties test, with few viable alternatives.  Three of the Sources (4, 6, and 10) display relatively constant X-ray luminosities over the 19 epochs, but again noting that only Source~6 has a secure identification (late-type subgiant, probably an RS~CVn).  The other seven are more variable, in some cases (2, 5, and 8) extremely so, with flare decays and flare outbursts dominating a few of the exposures, and weak or null detections in the others.  The {\em Gaia}\/ counterparts of the three most variable stars are consistent with M-type dwarfs, a class iconic for its flaring.  The four remaining objects are intermediate in variability: Source~1 is a moderately active G dwarf; Sources 3 and 9 apparently are active late-type giants; and Source 7 possibly is an F-type short-period binary.

\section{CONCLUSIONS}

The 10-year sequence of nineteen 10~ks {\em Chandra}\/ exposures of the $\alpha$~Cen field is a unique resource for evaluating the types of high-energy objects that might be found in a more-or-less arbitrary area near the Galactic plane (thereby avoiding extragalactic interlopers), and their behavior over short, intermediate, and long time scales.  The present study describes a synergistic approach to the X-ray source characterization problem, namely matching entries from the extensive Data Release 2 of the {\em Gaia}\/ astrometry mission to identify viable optical counterparts.  The high precision, sub-arcsecond source centroids delivered by {\em Chandra}\/ provide a stringent test of possible matches to the even higher precision (sub-mas) {\em Gaia}\/ positions.  The uniform broad-band $G$-magnitudes and $(G_{BP}-G_{RP})$ colors, combined with the deep limits of the {\em Gaia}\/ photometric survey ($G\sim 20$~mag), and the depth and precision of the parallaxes, allow a potential optical counterpart to be accurately placed in the observational H--R diagram, which then normally tightly constrains the nature of the object, especially if a single star (or optically dominant component of a binary).  Given the distance, the X-ray luminosity can be derived; while the bolometric flux can be calculated from the $G$-magnitude combined with a color-dependent bolometric correction.  The $L_{\rm X}$ and $L_{\rm X}/L_{\rm bol}$ ratios are two further metrics to either confirm an identification, or flag unexpected behavior.  This characterization scheme was successful for most of the objects-of-interest in the $\alpha$~Cen 10~ks sample.  At the same time, the approach revealed a few cases of moderately to highly inconsistent optical counterparts.  Future {\em Gaia}\/ data releases might resolve the discrepancies, for example by revealing even fainter, but spatially better-matched optical candidates.  At the moment, however, the true natures of the anomalous sources remain in doubt.

In the larger context, the clear message from the time-domain study is that the majority of high-energy stellar sources in the deep exposure of the $\alpha$~Cen field, and probably in general, are variable objects, with a high percentage of flaring M-type dwarfs.  The 190~ks HRC-I cumulative exposure has a limiting $5\,\sigma$ detection threshold of about $5{\times}10^{-15}$ erg cm$^{-2}$ s$^{-1}$.  For an active late-M dwarf, radiating at the saturation limit in X-rays, the limiting $G$-magnitude would be about +19~mag, close to the {\em Gaia}\/ DR2 photometric horizon; the corresponding distance would be a few hundred pc.  Given the large volume probed by the deep {\em Chandra}\/ exposure, it's no wonder that a large fraction of the detected objects are hyperactive M-types (red dwarfs are so common in the Galaxy that even the small percentage that are at the saturation limit become themselves a populous group).  

The high degree of stellar variability seen in the $\alpha$~Cen field has an important implication for soft X-ray surveys.  Namely, a census of an area of sky by overlapping single-epoch snapshots is likely to miss many of the transient sources, arguably among the most interesting objects.  Even for relatively normal coronal sources, there is a temporal bias introduced by magnetic cycles (e.g., Fig.~2 for $\alpha$~Cen B): the detectability of sunlike low-activity stars varies on a roughly decadal timescale according to the primary effect of higher mean coronal intensities near spot maximum, as well as lesser influences of enhanced transient flaring and spectral hardening on top of that.  Thus, a proper census of coronal sources over broad areas of the sky likely would require sampling each region repeatedly on a variety of timescales, to have a good chance of capturing the flaring objects as well as those experiencing peaks in long-term activity cycles (a coronal low state could linger for several years before brightening up by factors of several to ten into a high state).  

\section{FOR THE FUTURE}

This brings us back to the up-coming eROSITA survey mission, which is designed to scan the whole sky, every six months, for as many as four years, possibly longer.  The X-ray detection flux limits will be somewhat higher than for the {\em Chandra}\/ $\alpha$~Cen field (about $1{\times}10^{-14}$ erg cm$^{-2}$ s$^{-1}$ for eight 6-month all-sky scans: Predehl \& eROSITA Team 2013), but that means that the vast majority of the stellar content of the survey will appear in {\em Gaia}\/ DR2 (and its successors) with minimal ambiguities.  Not only will eROSITA transcend the limitations of previous stellar (and extragalactic) X-ray sky surveys, but also will provide opportunities for supporting measurements (especially in the ultraviolet) to advance our understanding of these pivotal, though still mysterious, high-energy coronal objects. Further, eROSITA surely will uncover, as here, numerous sources that do not fit in the current paradigm; potentially perhaps the most instructive cases in the long run.

\acknowledgments
This work was supported by grants from Smithsonian Astrophysical Observatory, based on observations from {\em Chandra}\/ X-ray Observatory, collected and processed at {\em Chandra}\/ X-ray Center, operated by SAO under NASA contract.  This study also made use of public databases hosted by {SIMBAD}, at {CDS}, Strasbourg, France; the Gator interface from NASA/IPAC (\url{https://irsa.ipac.caltech.edu/applications/Gator/}); and especially Data Release 2 of ESA's {\em Gaia}\/ mission (\url{https://www.cosmos.esa.int/gaia}) by {\em Gaia}\/ Data Processing and Analysis Consortium (\url{https://www.cosmos.esa.int/web/gaia/dpac/consortium}), funded by national institutions participating in the {\em Gaia}\/ Multilateral Agreement.   



\begin{thebibliography}{}

\bibitem[Argiroffi et al.(2016)]{2016A&A...589A.113A} Argiroffi, C., Caramazza, M., Micela, G., et al.\ 2016, \aap, 589, A113 
\bibitem[Ayres et al.(1998)]{1998ApJ...496..428A} Ayres, T.~R., Simon, T., Stern, R.~A., et al.\ 1998, \apj, 496, 428
\bibitem[Ayres(2004)]{2004ApJ...608..957A} Ayres, T.~R.\ 2004, \apj, 608, 957
\bibitem[Ayres et al.(2006)]{2006ApJS..165..618A} Ayres, T.~R., Plymate, C., \& Keller, C.~U.\ 2006, \apjs, 165, 618 
\bibitem[Ayres(2014)]{2014AJ....147...59A} Ayres, T.~R.\ 2014, \aj, 147, 59 
\bibitem[Ayres(2017)]{2017ApJ...837...14A} Ayres, T.~R.\ 2017, \apj, 837, 14
\bibitem[Ayres(2018)]{2018RNAAS...2a..17A} Ayres, T.~R.\ 2018, RNAAS, 2, 17 
\bibitem[Casagrande \& VandenBerg(2018)]{2018MNRAS.479L.102C} Casagrande, L., \& VandenBerg, D.~A.\ 2018, \mnras, 479, L102 
\bibitem[Forveille et al.(2018)]{2018A&A...616E...1F} Forveille, T., Kotak, R., Shore, S., \& Tolstoy, E.\ 2018, \aap, 616, E1 
\bibitem[Gaia Collaboration et al.(2018)]{2018A&A...616A..10G} Gaia Collaboration, Babusiaux, C., van Leeuwen, F., et al.\ 2018, \aap, 616, A10 
\bibitem[Giacconi et al.(2002)]{2002ApJS..139..369G} Giacconi, R., Zirm, A., Wang, J., et al.\ 2002, \apjs, 139, 369 
\bibitem[Lallement et al.(2018)]{2018A&A...616A.132L} Lallement, R., Capitanio, L., Ruiz-Dern, L., et al.\ 2018, \aap, 616, A132
\bibitem[Maehara et al.(2012)]{2012Natur.485..478M} Maehara, H., Shibayama, T., Notsu, S., et al.\ 2012, \nat, 485, 478 
\bibitem[Monet et al.(2003)]{2003AJ....125..984M} {Monet, D., Levine, S., Canzian, B., et al.\ 2003, \aj, 125, 984}
\bibitem[Osten et al.(2003)]{2003ApJ...582.1073O} Osten, R.~A., Ayres, T.~R., Brown, A., Linsky, J.~L., \& Krishnamurthi, A.\ 2003, \apj, 582, 1073
\bibitem[Osten et al.(2010)]{2010ApJ...721..785O} Osten, R.~A., Godet, O., Drake, S., et al.\ 2010, \apj, 721, 785   
\bibitem[Predehl \& eROSITA Team(2013)]{2013MmSAI..84..770P} Predehl, P., \& eROSITA Team 2013, \memsai, 84, 770 
\bibitem[Schmitt \& Liefke(2004)]{2004A&A...417..651S} Schmitt, J.~H.~M.~M., \& Liefke, C.\ 2004, \aap, 417, 651 
\bibitem[Sciortino et al.(2005)]{2005MmSAI..76..271S} Sciortino, S., Feigelson, E.~D., \& COUP Team 2005, \memsai, 76, 271 
\bibitem[Skrutskie et al.(2006)]{2006AJ....131.1163S} {Skrutskie, M.~F., Cutri, R.~M., Stiening, R., et al.\ 2006, \aj, 131, 1163}
\bibitem[Walter et al.(1980)]{1980ApJ...236..212W} Walter, F.~M., Cash, W., Charles, P.~A., \& Bowyer, C.~S.\ 1980, \apj, 236, 212
\bibitem[Wolk et al.(2005)]{2005ApJS..160..423W} Wolk, S.~J., Harnden, F.~R., Jr., Flaccomio, E., et al.\ 2005, \apjs, 160, 423
\bibitem[Wright et al.(2011)]{2011ApJ...743...48W} Wright, N.~J., Drake, J.~J., Mamajek, E.~E., \& Henry, G.~W.\ 2011, \apj, 743, 48


\end{thebibliography}
\end{document}